\documentclass{article}
\usepackage[tbtags]{amsmath}
\usepackage{amssymb}
\usepackage{amsthm}
\usepackage{array}
\usepackage{xy}

\usepackage[iso-8859-7]{inputenc}
\usepackage{graphicx}

\def\<{\left\langle}
\def\>{\right\rangle}

\title{\bfseries THERMALLY FAVOURABLE GAUGE MEDIATION \normalfont}
\author{Ioannis Dalianis and Zygmunt Lalak\footnote{e-mail addresses: Ioannis.Dalianis@fuw.edu.pl, Zygmunt.Lalak@fuw.edu.pl} \\
Institute of Theoretical Physics, Faculty of  Physics \\
University of Warsaw,
ul. Ho\.za 69, Warsaw, Poland}
\begin{document}

\date{}

\newcommand{\g}{\greektext} 
\newcommand{\e}{\latintext}
\maketitle

\abstract{We discuss the thermal evolution of the spurion and messenger fields of ordinary gauge mediation models taking into account the Standard Model degrees of freedom. It is shown that for thermalized messengers the metastable susy breaking vacuum becomes thermally selected provided that the susy breaking sector is sufficiently weakly coupled to messengers or to any other observable field.}

\section{Introduction} \normalsize

Gauge mediation is an attractive way of generating soft susy breaking in the Supersymmetric Standard Model \cite{Dine:1981za}-\cite{Dine:1993yw}. There exist viable models of gauge mediation, complete with detailed hidden sectors where susy is broken dynamically through strong dynamics; for a recent review see \cite{Kitano:2010fa}. Since the details of the hidden sector are often phenomenologically irrelevant, the hidden sector is parameterized by a singlet field $X$ which is a spurion of susy breaking and messengers $\phi$, $\bar{\phi}$ that through gauge interactions communicate susy breaking from the hidden sector to the Supersymmetric Standard Model fields. The most general renormalizable, gauge invariant and $R$-symmetric superpotential  is
\begin{equation} \label{W-OGM}
W=FX + (\lambda_{ij} X + m_{ij} )\phi_i \bar{\phi}_j.
\end{equation}
This theory can give either vanishing or non-vanishing gaugino masses at the leading order and a classification of the different cases can be found in \cite{Cheung:2007es}. It was shown in \cite{Komargodski:2009jf} that the gaugino masses are closely related to the vacuum structure of the theory. The formula for the gaugino masses at leading order in susy breaking is
\begin{equation} 
m_{\tilde{g}} \sim F^\dagger \frac{\partial}{\partial X} \log \det (\lambda_{ij} X +m_{ij})
\end{equation} 
and one can see that they vanish when $\det(\lambda X +m)=\det m$. In this case the origin $X=0$ is locally stable because there the scalar  messengers have positive squared masses. On the other hand, when $\det(\lambda X +m)$ depends on $X$, the gaugino masses are nonzero at the leading order. But there is a price to pay: there are no bare masses to protect all the messengers from becoming tachyonic for $|X|<X_{\text{min}}$, i.e. at the origin of field space. This implies the necessity the spurion field $X$ to be stabilized at a nonzero vacuum expectation value (vev) far from the origin. The superpotential (\ref{W-OGM}) doesn't determine the vev of $X$ which is, at tree level, a pseudo-modulus. It can get a potential from perturbative quantum corrections in the effective theory which lift the degeneracy. The Coleman-Weinberg potential usually stabilizes the pseudo-modulus at $X=0$ which implies that the potential runs off to infinity or to a susy vacuum at $X=0$, $\phi$, $\bar{\phi}\neq0$. 

Therefore, one has either to turn to models with locally stable origin and a mass hierarchy between sfermions and gauginos (ISS \cite{Intriligator:2006dd} and other direct mediation models fall to this category however, deformations of the ISS can evade this problem, \cite{Kitano:2006xg} is a first example) or to look for ways to stabilize the spurion at an $X\neq 0$ minimum. The former direction conflicts with a light Higgs necessary for the generation of the electroweak scale, except if one is ready to accept a more severe fine tuning in the Higgs sector. The later direction, fortunately, is not a blind siding. Gravitational effects and the need to cancel the cosmological constant in the phenomenologically acceptable vacuum can shift the susy breaking minimum at $X\neq0$ outside the tachyonic region \cite{Kitano:2006wz}. Also, it has been shown \cite{Shih:2007av} that when there are fields with $R$-charges $R\neq0,2$ the 1-loop corrections can create an $R$-symmetry breaking minimum at $X\neq0$.
Adding an explicit $R$-symmetry breaking mass term for messengers can stabilize the susy breaking minimum as well \cite{Murayama:2006yf}.

Despite the above positive results the theories (\ref{W-OGM}) with metastable vacua that give non-vanishing gaugino masses are cosmologically questioned. The thermal evolution of the hidden sector-messenger fields disfavours the selection of these susy breaking minima \cite{Craig:2006kx, Katz:2009gh, Dalianis:2010yk}. The free energy density minimizes at the origin of the field space and as the temperature decreases a phase transition towards the susy preserving vacuum takes place. On the contrary, vacua that give vanishing leading order gaugino masses are generally thermally preferred, see \cite{Abel:2006cr, Craig:2006kx, Fischler:2006xh} for the ISS model. 

The cosmological selection of these phenomenologically viable theories can be accomplished assuming a non-thermal evolution as in \cite{Ibe:2006rc} or even with thermalized messengers \cite{Dalianis:2010yk}. 
Nevertheless, whatever the proposed solution was, the spurion $X$ had to be in a particular way displaced at the time of reheating and obviously, the exact value of displacement is highly model dependent. 

In this Letter, we show that the Supersymmetric Standard Model (SSM) degrees of freedom can change the thermal history of the gauge mediation models in the limit of small coupling between the susy breaking and the messenger sector. We continue the discussion of \cite{Dalianis:2010yk} taking into account the  SSM fields explicitly. We show that when $\lambda \ll 1$ the metastable susy breaking vacuum can be thermally selected. Thermal selection means that the messengers are thermalized, i.e. the reheating temperature is large enough. The metastable vacuum is realized without the domination of the energy density of the universe by the spurion and hence, without a late entropy production. 

\section{Metastable Gauge Mediated Susy Breaking} \normalsize

The minimal model of ordinary gauge mediation is 
\begin{equation} \label{mm} 
W=FX-\lambda X \phi\bar{\phi}
\end{equation}
where $X$ is a standard model gauge singlet field and $\phi$, $\bar{\phi}$ messenger fields carrying Standard Model quantum numbers.
The scalar potential in the global limit reads 
\begin{equation}  
\, \,\,\,\,\,\,\,\,\,\,\,\,\,\,\,\,\,\,\,\,\,\,\,\,\,\,\,\,\,\,\,\,\,\,\,\,\,\,\,\,\,\,\, V_F = |F|^2 + |\lambda|^2 |X|^2 \left(|\phi|^2+|\bar{\phi}|^2\right) - (F^\dagger \lambda \phi\bar{\phi}+ \text{h.c.} ) + |\lambda|^2 |\phi|^2 |\bar{\phi}|^2 \,\,\,\,\,\,\,\,\,\,\,\,\,\,\,\,\,\,\,\,\,\,\,\,\,\,\,\,\,\,\,\,\,\,\,\,\,\,\,\,\,\,\,\,
\end{equation}
for canonical K\"ahler. The  $\lambda$ and $F$ can be considered real after a phase rotation. The $X=0,\,\phi\bar{\phi}=F/ \lambda$ is a supersymmetric flat direction. The $\phi=\bar{\phi}=0, X$ is the susy breaking flat direction with $X$ not determined at tree level.
An $R$-symmetric extension of the minimal model (\ref{mm}) is to include an extra set of messengers plus a mass parameter:
\begin{equation} \label{mm2}
W=FX+\lambda X \phi_1 \bar{\phi}_1+\lambda X \phi_2 \bar{\phi}_2+m \phi_1 \bar{\phi}_2.
\end{equation}
The directions $\bar{\phi}_1$ and $\phi_2$ in the field space are not protected by the mass term. The area about the origin $|X|^2<F/\lambda$ is tachyonic for both models (\ref{mm}) and (\ref{mm2}).

The degeneracy along the $X$-direction can be lifted. The interaction term $\lambda X \phi\bar{\phi}$ induces at one-loop level a correction to the K\"ahler potential $\delta K  \simeq -(\lambda^2/16\pi^2) |X|^2\log(|X|^2/M^2)$ which attracts $X$ to the origin. In addition, the initial K\"ahler for the spurion may be non-canonical and take the form
\begin{equation} \label{K}
K=X^\dagger X -\frac{(X^\dagger X)^2}{\Lambda^2}
\end{equation}
with a cut-off scale $\Lambda$. For $|X|<\Lambda$ the potential scales like $V \sim |X|^2 F^2 /\Lambda^2$. Above that scale another (microscopic) theory takes over. A simple example is an O'Raifeartaigh type superpotential $W=m_{o}\chi_1\chi_2+k_o X\chi_1^2+FX$. For $\sqrt{F}\ll m_o$ the O'Raifeartaigh fields $\chi_1$ and $\chi_2$ are integrated out in and the effective superpotential is $W_{\text{low}}=FX+messengers$. The presence of the raifeartons is encoded in the K\"ahler potential which includes the one-loop contribution from $\chi$ fields 
and at low energies is effectively described by ($\ref{K}$) with $\Lambda^2 \sim m^2_o/k^4_o$. Another possibility is that $X$ is a composite particle 
which forms a bound state below the scale $\Lambda$.

The next question concerns the expectation value of the pseudo-modulus spurion. Obviously it has to be stabilized at $|X|>X_\text{min}$. Generally, this can happen thanks to gravity. Adding to the superpotential a dimensionful constant $c$ in order to cancel the cosmological constant at the susy breaking vacuum and for the K\"ahler (\ref{K}) the minimum is at \cite{Kitano:2006wz}
\begin{equation}
\left\langle X\right\rangle\simeq \frac{\sqrt{3}\Lambda^2}{6M_P}.
\end{equation}
There is also a way to give a vev to $X$ for the superpotential ($\ref{mm}$) even in the global limit. Assuming a K\"ahler potential 
\begin{equation} \label{6}
K=X^\dagger X + \frac{(X^\dagger X)^2}{\Lambda^2_1}-\frac{(X^\dagger X)^3}{\Lambda^4_2} 
\end{equation}
for $\Lambda^{3/2}_1/M^{1/2}_P<\Lambda_2<\Lambda_1$, the spurion is stabilized at $\left\langle X\right\rangle= \Lambda^2_2/\Lambda_1$, where $U(1)_R$ is spontaneously broken. 

For canonical K\"ahler the Coleman-Weinberg correction can give an $\left\langle X\right\rangle\neq0$, breaking also spontaneously the $R$-symmetry, if there are exotic messenger $R$-charges  \cite{Shih:2007av}. In particular for the superpotential ($\ref{mm2}$) there is a minimum at $\left\langle X \right\rangle \simeq 0.3 m/\lambda$. We note that the minimal model (\ref{mm}) cannot exhibit such a behaviour because there is no field with charge $R\neq 0, 2$.

Another simple solution to the problem of the spurion stabilization is to add to the superpotential (\ref{mm}) an explicit $U(1)_R$ violating mass term $M\phi \bar{\phi}$ for the messengers \cite{Murayama:2006yf, Murayama:2007fe}
\begin{equation}
W=FX-\lambda X \phi \bar{\phi}-M\phi \bar{\phi}.
\end{equation}
This relegates the susy vacua to $X\neq 0$ and a K\"ahler of the form (\ref{K}) can stabilize safely the spurion at the origin. This model has similarities with the gravitational stabilization. Here, instead of the constant $c$ it is the mass $M$ that violates the $R$-symmetry. After the transformation $X \rightarrow \tilde{X}=X+M/\lambda$ the superpotential and the K\"ahler metric read respectively $W=F\tilde{X}-\lambda \tilde{X}\phi \bar{\phi}- FM/\lambda$ and $K_{\tilde{X}^\dagger\tilde{X}}=1-4\left(|\tilde{X}|^2-(\tilde{X}+\tilde{X}^\dagger)M/\lambda+(M/\lambda)^2\right)/\Lambda^2$. This will result in a term linear in $\tilde{X}$ that shifts the minimum of the susy breaking vacuum to $\left\langle \tilde{X} \right\rangle=M/\lambda$. The susy preserving is at $\tilde{X}=0,\, \phi\bar{\phi}=F/\lambda$.
\\
\\
The fact that the susy breaking vacuum is a local minimum in the field space, with an unstable origin, makes these theories cosmologically doubtful.
The messengers carrying $SU(3)\times SU(2)\times U(1)$ quantum numbers get thermalized and, also, induce thermal masses on the spurion $X$ \cite{Dalianis:2010yk}. The unstable origin becomes the minimum of the finite temperature effective potential since the thermal masses compensate the tachyonic ones. For coupling $\lambda$ of order ${\cal O} (1)$ there is a second order phase transition towards the susy preserving vacua.

However, as we will demonstrate in the next section,  in the limit $\lambda \ll 1$ the thermal evolution radically changes and the selection of the metastable vacuum can take place naturally. The small coupling is necessary in order the thermal mass of the spurion to stay small and hence, the metastable vacuum to emerge from the thermal corrections at high temperatures. On the other hand, while $\lambda$ decreases, the messenger thermal masses cannot become arbitrary small thanks to the SSM degrees of freedom.  Thus, the messenger tachyonic masses are 'covered' by the thermal ones until lower temperatures.  Asking for a particularly small coupling between messengers and the spurion prompts us to check whether other interactions could alter this picture. Actually, only if the exact interactions of messengers and spurion fields are known one can trace the thermal evolution of the system. Below, we will briefly summarize some extensions of the minimal interactions (\ref{mm}) of $X$, $\phi$ and $\bar{\phi}$ with SSM fields.
 
Firstly, one can assume that there is a mixing of the messenger fields with ordinary fields. This could enhance further the thermal effects.
The messenger superfields $\bar{\phi}$ have the same quantum numbers as the ordinary, visible $\bar{d}$ superfields. The difference is that the former have couplings of the form $X \phi\bar{\phi}$ whereas, the later have Yukawa couplings, $QH_D \bar{d}$.  Thus, one can consider a simple modification that takes place in the Yukawa sector. In particular a messenger-matter mixing \cite{Dine:1996xk}
\begin{equation} \label{mixing}
H_D L_iY^l_{ij}\bar{e}_j+H_D Q_iY^d_{ij}\bar{d}_j
\end{equation}
with each of $L_i$ and $\bar{d}_i$ refers to the four objects with the same quantum numbers. The convention is that the $L_4$ and $\bar{d}_4$ are a linear combination of fields which couple to the spurion $X$. $Y^l$ is a $4\times 3$ matrix while $Y^{d}$ is a $3\times 4$ matrix, and the $Y^l_{4i}$ and $Y^d_{i4}$ are the "exotic" Yukawa couplings. The above messenger-matter mixing, if present can also contribute to the thermal mass squared of messengers with an additional $(|Y^{l}_{4i}|^2+|Y^{d}_{i4}|^2)T^2$ term in the effective potential. However, this mixing results in non-universal contributions to scalar masses and FCNC constraints 
the exotic Yukawa couplings to be weaker than the ordinary Yukawa couplings. 

Another possibility is that the messengers couple to the Higgs superfields in the superpotential  
\begin{equation} \label{Hq}
W=kH_U\phi_1\phi_2+\bar{k}H_D\bar{\phi}_1\bar{\phi}_2.
\end{equation}
This coupling was proposed in \cite{Dvali:1996cu} in order to generate a $\mu$-term at one-loop level. For $k={\cal O}(1)$ this coupling can induce a significant thermal mass on messengers. 

On the other hand the gauge singlet $X$ may have direct couplings to SSM Higgs superfields  
\begin{equation} \label{HX}
W \supset \epsilon X H_U H_D  
\end{equation}
with a small coefficient $\epsilon$. This interaction was introduced in order to generate at tree level a $B_\mu$-term for the Higgs sector \cite{Dine:1995ag}. For low energy phenomenological reasons it has to be negligible small. If $\epsilon< \lambda $ then it is negligible in the finite temperature effective potential as well.

To sum up, the couplings in the case of (\ref{mixing}) are negligible, whereas the (\ref{Hq}) and (\ref{HX}) may be important and could modify the critical temperature of the phase transitions. In the next section we will present the thermal evolution of the spurion and messengers taking into account only the gauge vector fields which by definition are present and probably account for the most significant thermal contributions. We will comment on the possible effects of (\ref{Hq}) and (\ref{HX}) combined with the cosmological constraints in the conclusions.

\section{Thermal Evolution} \normalsize
If one neglects the SSM degrees of freedom, the thermalized system of fields evolves towards the susy preserving vacua \cite{Dalianis:2010yk}\footnote{Except if the system of fields, for $\lambda \ll 1$, is trapped close to the metastable vacuum after inflation \cite{Dalianis:2010yk}.}.
This makes perfect sense. Having only one coupling $\lambda$ to the thermal plasma messengers and spurion are equally influenced by the thermal equilibrium. The messengers, having tree level masses, are heavier than the spurion which receives a mass of quantum origin (either due to non-minimal K\"ahler or to Coleman-Weinberg corrections from the interaction with messengers). Therefore, higher temperatures are required to overwhelm the messenger tree level masses rather than the small 'quantum' mass of $X$.

Including the Standard Model gauge bosons introduces an extra contribution $gT$ to the messenger thermal masses but \itshape not \normalfont to the thermal mass of the spurion. Decreasing the coupling $\lambda$ the spurion thermal mass, $\lambda T$, is suppressed while the messengers' remains $gT$. Therefore, there is a threshold value of the coupling, $\lambda_\text{max}$, that below this value the phase transitions get inversed: the transition to the metastable susy breaking vacuum precedes the transition to the supersymmetric one. 

At finite temperature the fields that interact with the thermal plasma are no longer in their vacuum state. The occupation numbers $n_\bold{k}$ are given by the Bose-Einstein formula. The temperature dependent 1-loop effective potential is of the form \cite{Mukhanov:2005sc, Quiros:1999jp}  $V^T_1 \sim T^4 \int dx x^2 \ln \left(1 \pm \text{exp}\left(-\sqrt{x^2+M^2_i/T^2}\right)\right)$ where $M^2_i$ is an eigenvalue of the mass squared matrices. In the high temperature limit where $T$ is much greater than the mass eigenvalues the scalar potential reads 
\begin{equation} \label{hT}
\bar{V}^T_1(\phi_c)\simeq -\frac{\pi^2 T^4}{90} \left(N_B+\frac{7}{8}N_F\right)+\frac{T^2}{24}\left[\sum_{i}(M^2_S)_i+3\sum_{a}(M^2_V)_a+ \sum_{r}(M_F)^2_r\right]
\end{equation} 
where we have omitted the negligible terms linear in temperature. We are interested in how the thermal effects change the shape of the potential and in particular in the position of the high temperature minima and how they evolve as the temperature decreases. 
\begin{figure} 
\textbf{\,\,\,\, GRAVITATIONAL STABILIZATION \,\,\,\,\,\,\,\,\,\,\,\,\,\,\,\,\,\,\,\;\;\;\;\;\;\;\;\;\; MESSENGER MASS}
\centering
\begin{tabular}{cc}

{(a)} \includegraphics [scale=.85, angle=0]{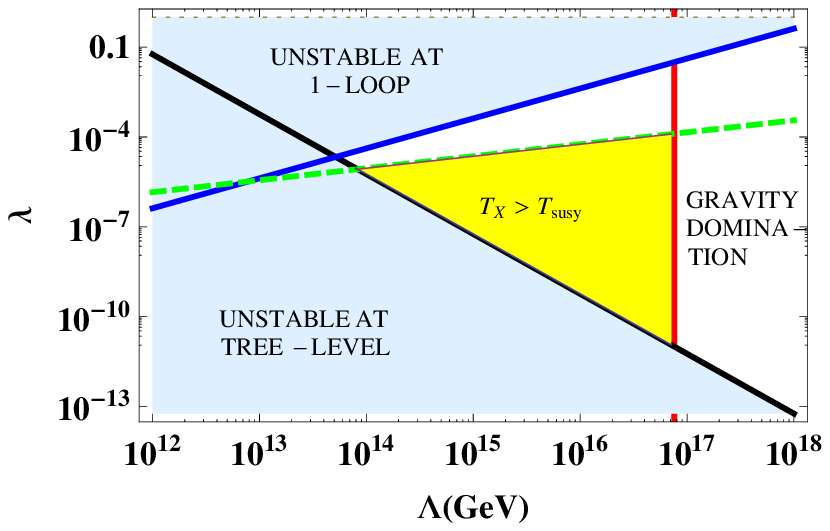} &
{(b)} \includegraphics [scale=.85, angle=0]{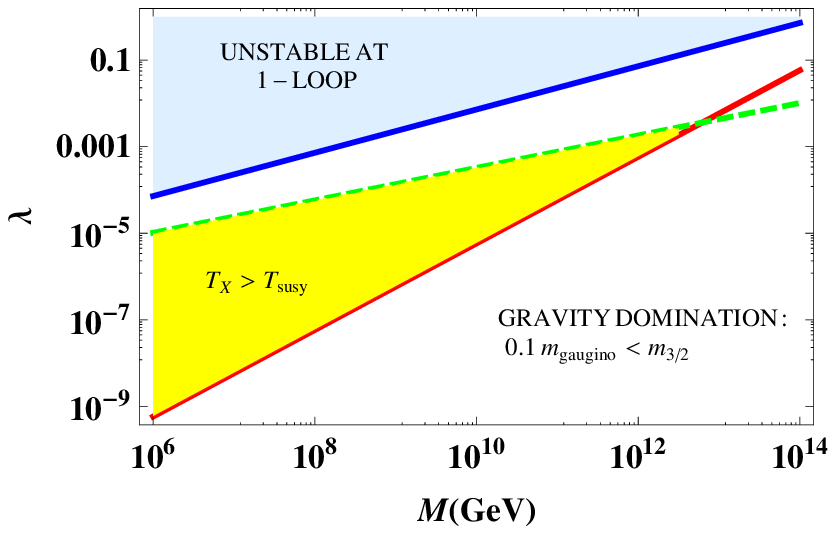}  \\
\end{tabular}
\caption{\small{The figures show the region of parameter space where the supersymmetry breaking minimum is metastable (white and yellow region) for the case of (a) gravitational stabilization and (b) messenger mass $M\phi\bar{\phi}$. In the yellow region, below the green dashed line, the thermal selection of the metastable vacuum is realized. The red line separates gauge mediation, $m_{3/2}<0.1 m_{\tilde{g}}$, from gravity mediation. In the panel (b) the K\"ahler correction scale is fixed at $\Lambda=2.4\times 10^{15}$ GeV.}}
\end{figure}
In principle we should include the $D$-terms in the scalar potential i.e. $V=V_F+V_D$.
The $D$-term contribution to the sfermion masses can vanish if one imposes the "messenger parity" proposed in \cite{Dimopoulos:1996ig} -but not for the superpotential (\ref{mm2}).
Although the $D$-terms may be problematic for the low energy phenomenology they do not change essentially the thermal evolution of the fields. 

The messenger superfields $\phi+\bar{\phi}$ can be decomposed into colour triplets $q+\bar{q}$ and weak doublets $\ell+\bar{\ell}$ which couple to the spurion like 
\begin{equation}
\lambda_q X q \bar{q}+\lambda_\ell X \ell \bar{\ell}\, ,
\end{equation}
where we considered the general case of different $\lambda_\ell$ and $\lambda_q$. In the case of unification, e.g. when the messengers transform in the $\bold{5}+\bold{\bar{5}}$ representations of $SU(5)$ one has $\lambda_\ell=\lambda_q=\lambda$. We explicitly write it in this way instead of the compact form (\ref{mm}) because the doublets $\ell +\bar{\ell}$ are coupled to the thermal plasma weaker than the colour triplets $q+\bar{q}$. 

At high temperatures the thermal masses squared compensate the negative ones and the effective minimum lies at the origin\footnote {Except if there are $U(1)_R$ violating terms like the constant $c$ or messenger mass terms of the form $M\phi\bar{\phi}$ in the superpotential. Then, the thermal minimum is at $X\neq 0$.} of field space. 
Apart from the spurion\footnote{For weak coupling $\lambda_{\ell,\,q}$ the spurion is actually out of equilibrium \cite{Dalianis:2010yk}.} and the self-coupling of the messengers, the gauge bosons induce thermal masses for $\ell +\bar{\ell}$ and $q+\bar{q}$. The interaction between the observable gauge bosons and the messengers is identified in the kinetic terms. For the scalar messenger fields the covariant derivatives read
\begin{equation}
D_\mu \ell= \partial_\mu \ell - i g W^a_\mu \frac{\tau ^a}{2} \ell - i y_\ell \frac{g'}{2}B_\mu \ell
\end{equation}
and
\begin{equation}
D_\mu q= \partial_\mu q - i g_s G^a_\mu \frac{\tilde{\lambda}_{a}}{2}q - i y_q \frac{g'}{2}B_\mu q
\end{equation} 
with $y_{\ell,q} =-1,\,-2/3$ and $y_{\bar{\ell},\bar{q}}=1, \, 2/3$. The triplets $q +\bar{q}$ couple to the thermal plasma mainly via the strong gauge coupling $g_s$. 
The doublets $\ell+\bar{\ell}$ couple with $g$ and $g'/2$ of $SU(2)$ and $U_Y(1)$ respectively.  At high energy $g_s$ runs weaker and $g$, $g'$ stronger. For the temperatures discussed here we consider the approximate values $g^2_s \sim 4\pi/17$, $g^2\sim 4\pi/28$ and $g'^2 \sim 4\pi/43$.

In the vicinity of the origin i.e., $q=\bar{q}=X=0$, the relevant terms in the temperature corrected scalar potential for the doublets $\ell+\bar{\ell}$ are
\begin{equation}
V \supset -\lambda_\ell F\left(\ell \bar{\ell}+\text{h.c.}\right) + \lambda^2_\ell |\ell \bar{\ell}|^2+ \frac{T^2}{24}\left[\left(6\lambda^2_\ell+\frac{9}{2}g^2+\frac{3}{2}g'^2\right)\left(|\ell|^2+|\bar{\ell}|^2\right)\right].
\end{equation}
In the ${\cal O}(T^2)$ part of the potential the contribution of the fermions and the spurion to the thermal masses of $\ell$,$\bar{\ell}$ has been taken into account. A decoupled spurion would decrease the coefficient in front of the $\lambda_\ell$ by a factor of 3. We recall that in the above expression the $\ell$,$\bar{\ell}$ refer to the thermal average values since the potential is corrected by the temperature dependent part. 
The critical temperature, i.e. the temperature when the mass squared at the origin turns from positive to negative can be seen easier after a diagonalization of the mass matrix. We rotate the fields to $L_1= (\bar{\ell}^\dagger+\ell)/\sqrt{2}$ and $L_2=(\bar{\ell}-\ell^\dagger)/\sqrt{2}$ and the mass terms in the potential are transformed to
\begin{equation}
V \supset -\lambda_\ell F \left(|L_1|^2-|L_2|^2\right) + \frac{T^2}{24}\left(6\lambda^2_\ell+\frac{9}{2}g^2+\frac{3}{2}g'^2\right) \left(|L_1|^2+|L_1|^2\right). 
\end{equation}
The direction $L_1$ becomes tachyonic at temperature 
\begin{equation} \label {Tcr1}
T^\ell_{\text{susy}} = 4\sqrt{\frac{\lambda_\ell F}{3g^2+g'^2+4\lambda^2_\ell}} \, .
\end{equation}
At this temperature, to a good approximation, a second order phase transition towards the supersymmetric vacuum takes place. In the case that $\lambda_\ell \sim 1$ so, $\lambda_\ell >g$ the critical temperature will be 
\begin{equation}
T^\ell_\text{susy} \simeq 2\sqrt{\frac{F}{\lambda_\ell}}, \,\,\,\,\,\,\,\,\,\,\,\,\, \text{for $\lambda_\ell \sim 1$}.
\end{equation}
We see that a small coupling, $\lambda_\ell \ll g'$, between the spurion and the messenger fields decreases the critical temperature $ \sqrt{(3g^2+g'^2)/4\lambda^2_\ell}$ times compared to the case of negligible gauge bosons contribution.  The values of gauge couplings, as mentioned above, are $3g^2+g'^2 \simeq 1.64$ hence, for $\lambda_\ell \ll 1$ the decrease can be significant.

In the case that the $D$-terms don't vanish the relevant potential reads
\begin{equation}
V=V_F+\frac{1}{2}g^2\left(\ell^\dagger\frac{\vec{\tau}}{2}\ell+\bar{\ell}^\dagger\frac{\vec{\tau}}{2}\bar{\ell}\right)^2+\frac{1}{2}\left(\frac{g'^2}{2}\right)^2\left(\ell^\dagger \ell- \bar{\ell}^\dagger \bar{\ell} \right)^2.
\end{equation}
The effective (thermal) masses of $\ell$, $\bar{\ell}$ will obtain an extra contribution, but it is of the same order of magnitude as the previous one and the critical temperature is not essentially changed.

Following the same steps for the triplets $q+\bar{q}$ it is straightforward one to see that the critical temperature in this case is
\begin{equation}
T^q_\text{susy} = 4\sqrt{\frac{\lambda_q F}{8g^2_s+(4/9)g'^2+4\lambda^2_q}}.
\end{equation}
For weak couplings, $\lambda_q$,$\lambda_\ell \ll 1$, the critical temperature for the triplets $q+\bar{q}$ is lower than the critical temperature ($\ref{Tcr1}$) for the doublets $\ell+\bar{\ell}$ provided that $\lambda_q/\lambda_\ell<(8g^2_s+(4/9)g'^2)/(3g^2+g'^2)$. Pluging in the values of the running gauge couplings for tempereatures of the order $T\sim 10^{9} $ GeV the previous condition reads $\lambda_q/\lambda_\ell < 5$. Hence, considering $\lambda_q\sim\lambda_\ell$, the first (larger) critical temperature for the transition to the susy vacua is the one for the doublets. Hereafter we will assume the (\ref{Tcr1}) as the critical temperature for the system of fields. 

The metastable susy breaking vacua appear at temperature $T_X$. The exact value depends on the way the spurion $X$ is stabilized. We consider separately the cases of stabilization with and without gravity.

\subsection{Gravitational Stabilization}
\itshape Gravitational Gauge Mediation \normalfont \\
A minimal model is the one described in \cite{Kitano:2006wz} with $W=FX-\lambda X\phi \bar{\phi}+c$ and $K=|X|^2-|X|^4/\Lambda^2$. At zero temperature the origin is unstable in the direction of messengers for $|X|<\sqrt{F/\lambda_{\ell,\, q}}$ ; at temperatures $T>T_\text{susy}$ messengers thermal masses overtake the tachyonic ones. The spurion receives a thermal mass of order $\lambda_\ell T$ and $ \lambda_q T$ from doublets and triplets that stabilize $X$ close to zero. As the temperature decreases the thermal effects weaken and the minimum in the $X$-direction shifts towards the zero temperature value. The moment that it exits the (would-be at $T_\text{susy}$) tachyonic region, i.e. $X>\sqrt{F/\lambda_{\ell,\,q}}$, 
the metastable vacuum forms \cite{Dalianis:2010yk}. This takes place at temperature squared\footnote{For temperatures $T > T_\text{susy}$ there is single global minimum of the finite temperature effective potential. There are no tachyonic directions. In the case that $T_X > T_\text{susy}$ at the temperature $T_X$ the "would-be metastable" minimum forms; hence, initially the minimum in the $X$-direction is global and at $T_\text{susy}$ it becomes local i.e. metastable, but it never becomes unstable. It would become unstable only if $T_X<T_\text{susy}$.}
\begin{equation}
{T^\ell_X}^2 \simeq 8  \frac{c \sqrt{F}}{(2\lambda^2_\ell+3\lambda^2_q) M^2_P}\sqrt{\lambda_\ell}\,\, , \,\,\,\,\,\,\,\,\,\,\,\,\,\,\,\, {T^q_X}^2 \simeq  8\frac{c \sqrt{F}}{(2\lambda^2_\ell+3\lambda^2_q) M^2_P}\sqrt{\lambda_q}
\end{equation}
for doublets and triplets respectively. Considering $\lambda_\ell \simeq \lambda_q =\lambda$ the $T_X$ temperature reads
\begin{equation} \label{T_X}
T^2_X \simeq \frac{8}{5} \frac{c}{\lambda M^2_P}\sqrt{\frac{F}{\lambda}}
\end{equation}
with $c=FM_P/\sqrt{3}= m_{3/2}M^2_P$ for vanishing cosmological constant in the metastable vacuum.  
The $T_X$ can be larger than $T_\text{susy}$ for small coupling $\lambda$, namely
\begin{equation}
\lambda< \left(\frac{3g^2+g'^2}{10}\frac{\sqrt{F}}{\sqrt{3}M_P}\right)^{2/5} \simeq \left(0.16\frac{\sqrt{F}}{\sqrt{3}M_P}\right)^{2/5}.
\end{equation} 
To present an example for $\sqrt{F} =2.4\times 10^9$ GeV  ($m_{3/2} \simeq 1$ GeV) the coupling has to be lower than $\lambda<1.0 \times 10^{-4}$ and for $\sqrt{F} =2.4 \times 10^8$ GeV  ($m_{3/2}\simeq 10^{-2}$ GeV), $\lambda<3.9 \times 10^{-5}$. The scaling of temperatures $T_\text{susy}$ and $T_X$ is demonstrated in the figure 2. We recall that gaugino mass of the order of ${\cal O}(100)$ GeV relates the parameters $F$ and $\Lambda$ according to $F\simeq  10^{-14}\left\langle X \right\rangle M_P$. We also remind the reader that the coupling $\lambda$ cannot become arbitrary small or large  because of the constraints from the zero temperature stability conditions on the susy breaking vacuum: $10^{-14}(\Lambda/M_P)^{-2}<\lambda<\Lambda /M_P$ which are illustrated in figure 1.

In the case of gravitational stabilization, at $T_X$ there is no  phase transition; only a smooth shift of the vacuum to larger values. Hence, the system of fields lands at the metastable vacuum if the effective mass of the spurion $X$ is sufficiently larger than the Hubble scale. Following \cite{Linde:1996cx} we assume that when $M_X>30H$ the $X$ field tags along the position of the temperature dependent minimum and its oscillations are efficiently damped. In a radiation dominated phase $H=1.66 g^{1/2}_*T^2/M_P$ and $M_X \simeq \lambda T/\sqrt{2}$ as one finds from the finite temperature potential. This gives a lower limit on the ratio $\lambda/T>30\times 1.66\sqrt{2}g^{1/2}_* M^{-1}_P ={\cal O}(500) M^{-1}_P$. Otherwise, the spurion either stays frozen (when $M_X<H$) to its postinflationary value until lower temperatures or its oscillations about the metastable vacuum are not efficiently damped (when $H<M_X<30H$). For the values of coupling considered here, i.e. $ 10^{-8} \lesssim \lambda \lesssim 10^{-4}$, the limit on the ratio $\lambda/T$ is not problematic.

\subsection{Global Limit}
\itshape Higher Order K\"ahler Corrections \normalfont \\
For the case of 6th order corrected K\"ahler function (\ref{6})
the metastable vacuum survives in the global limit $M_P \rightarrow \infty$. We consider again $\lambda_\ell \simeq \lambda_q =\lambda$. The metastable vacua appear at \cite{Dalianis:2010yk}
\begin{equation} \label{gl-1}
T_X= \frac{4}{\sqrt{5}} \frac{1}{\lambda} \frac{F}{\Lambda_1}.
\end{equation}
neglecting gravity. Also here, we see that decreasing $\lambda$ increases the temperature $T_X$. For 
\begin{equation}
\lambda< \left( \frac{3g^2+g'^2}{5} \frac{F}{\Lambda^2_1}\right)^{1/3} \simeq \left(0.33\frac{F}{\Lambda^2_1}\right)^{1/3}
\end{equation}
$T_X$ is larger than $T_\text{susy}$ and there is a second order phase transition to the metastable vacuum. Supersymmetry and $U(1)_R$ break spontaneously. For $\sqrt{F} =2.4 \times 10^8$ GeV and $\Lambda_1=2.4\times 10^{14}$ GeV the coupling has to be smaller than  $\lambda< 6.9 \times 10^{-5}$ and for $\sqrt{F} =2.4 \times 10^{7.5}$ GeV and $\Lambda_1=2.4\times 10^{13}$ GeV we take $\lambda< 1.5 \times 10^{-4}$.
\\
\\
\itshape Messenger Mass \normalfont \\
In the case that there is an extra messenger mass term $M\phi\bar{\phi}$ in (\ref{mm}) with K\"ahler $K=|X|^2-|X|^4/\Lambda^2$ susy breaks down at $\phi=\bar{\phi}=X=0$ while the susy preserving minimum lies at $X=-M/\lambda, \phi\bar{\phi}=F/\lambda$, for  $\lambda_\ell \simeq \lambda_q =\lambda$. With a field transformation $X\rightarrow \tilde{X}=X+M/\lambda$ the vacua switch positions along the $X$-axis. The potential, then, has a form similar to the potential of the gravitational stabilization. Following the same steps, we find that the temperature at which the $X$-minimum exits the tachyonic region $|\tilde{X}|<\sqrt{F/\lambda}$ is 
\begin{equation} \label{TM}
T^2_{{X}} \simeq \frac{16}{5}\frac{FM}{\lambda^2\Lambda^2}\sqrt{\frac{F}{\lambda}}
\end{equation}
which is of course the same for $X$ and $\tilde{X}$. The temperature (\ref{TM}) is the analogue of (\ref{T_X}) with the correspondence $c/M^2_P \rightarrow 2 FM/(\lambda \Lambda^2)$. Fixing the gaugino mass to be of the order ${\cal O}(100)$ GeV, gives $F\simeq 10^{-14} M_P \left\langle \tilde{X} \right\rangle =10^{-14}M_PM/\lambda$. The main difference, is that here one has three parameters ($M,\Lambda, \lambda$) instead of two ($\Lambda, \lambda$) of the gravitational stabilization. The fact that the messengers have explicit mass $M$ that doesn't depend on the coupling $\lambda$ changes the behaviour of the critical temperature of the transition towards the susy vacua. Namely, from the last relation we take that $\lambda F \simeq 2.4 \times 10^{4}M$ GeV and the critical temperature (\ref{Tcr1}) reads
\begin{equation}
T^\ell_\text{susy} \simeq  10^{7} \, \text{GeV}\,\left(\frac{M}{2.4\times 10^{8}\,\text{GeV}}\right)^{1/2}\left(\frac{1}{3g^2+g'^2}\right)^{1/2}.
\end{equation}
For a given mass $M$ it has a fixed value. The $T_X$ is larger than $T^\ell_\text{susy}$ for 
\begin{equation}
\lambda< \left(0.41\times 10^{-2}\frac{M^2}{\Lambda^2}\right)^{1/4}\left(\frac{2.4 \times 10^{8}\,\text{GeV}}{M}\right)^{1/8}.
\end{equation}
Hence, for messenger mass $M=2.4\times 10^{8}$ GeV and cut-off scale $\Lambda=2.4\times 10^{15}$ GeV the transition to the susy breaking vacuum takes place first if $\lambda<8.0 \times 10^{-5}$; for $M=2.4\times 10^{6}$ GeV, $\Lambda=2.4\times 10^{15}$ GeV if $\lambda<1.4 \times 10^{-5}$, see figure 2. The coupling $\lambda$ cannot become arbitrary small because gravity contributions to the soft masses start to dominate, see figure 1. Note that, here, the vev of the spurion is $\left\langle \tilde{X}\right\rangle=M/\lambda$ and the coupling $\lambda$ is a free parameter. Hence, for fixed gaugino masses the gravitino mass will scale like 
\begin{equation}
m_{3/2} \simeq \left(\frac{M}{2.4\times 10^{8}\,\text{GeV}}\right)\left(\frac{10^{-6}}{\lambda}\right)\,\, \text{GeV}.
\end{equation} 

We recall here the zero temperature constraints that render the susy breaking vacuum metastable: $\lambda F < M^2$ and $\lambda^2<4\pi M/\Lambda$, see figure 1.
\\
\\
\itshape Canonical K\"ahler \normalfont \\
Finally, for the case of canonical K\"ahler, if there is e.g. a double set of messengers with $\delta W =m \phi_1 \bar {\phi}_2$ (\ref{mm2}) which have exotic $R$-charges both $U(1)_R$ and supersymmetry can break down spontaneously via a second order phase transition \cite{Cheung:2007es, Shih:2007av}. Although there are similarities with the case of K\"ahler corrected up to 6th order, here the stabilization of the spurion is basically different. It is due to the perturbative quantum correction coming from the interaction of the messengers with the $X$ field. The Coleman-Weinberg potential has a higher order dependence on the coupling i.e. $\lambda^2F^2$ to leading order in $F^2$. It is not straightforward to see from the effective potential which is not of a polynomial type the critical temperature analytically. The mass squared of the spurion scales like $\lambda^4 F^2/m^2$ and hence we expect that the $T_X$ takes the approximate form 
\begin{equation}
T_X \sim \frac{\lambda F}{m}.
\end{equation}
\begin{figure} 
\textbf{\,\,\,\,\,\,\, GRAVITATIONAL STABILIZATION \,\,\,\,\,\,\,\,\,\,\,\,\,\,\,\,\,\,\,\;\;\;\;\;\;\;\;\;\;\; MESSENGER MASS}
\\
\\
\centering
\begin{tabular}{cc}
{(a1)} \includegraphics [scale=.85, angle=0]{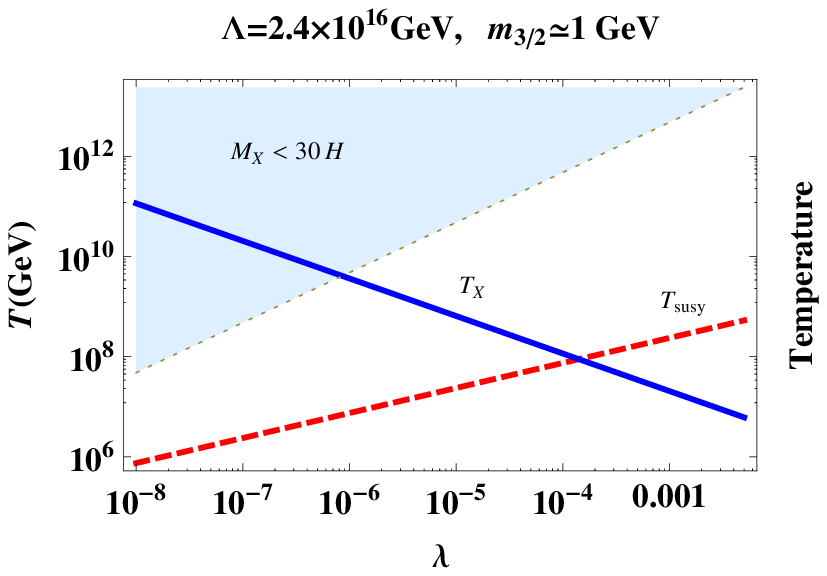} &
{(b1)} \includegraphics [scale=.85, angle=0]{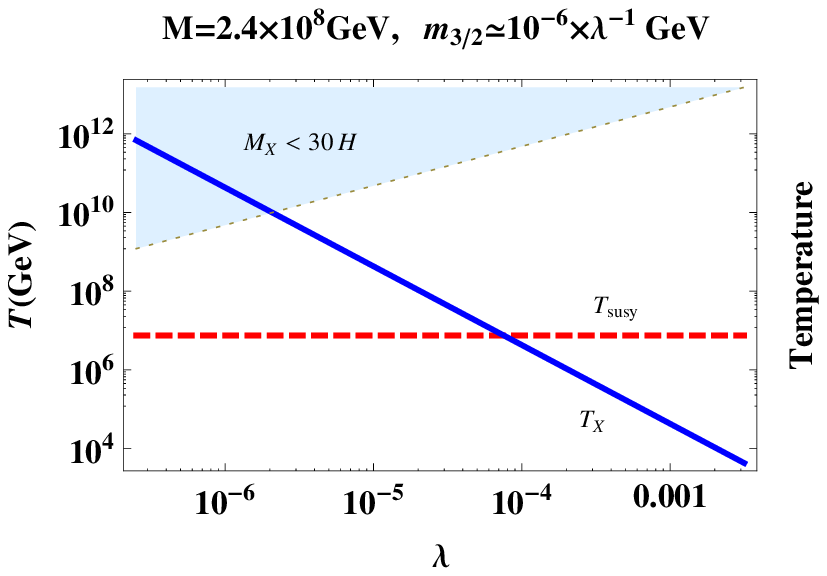}  \\
\end{tabular}
\begin{tabular}{cc}

{(a2)} \includegraphics [scale=.85, angle=0]{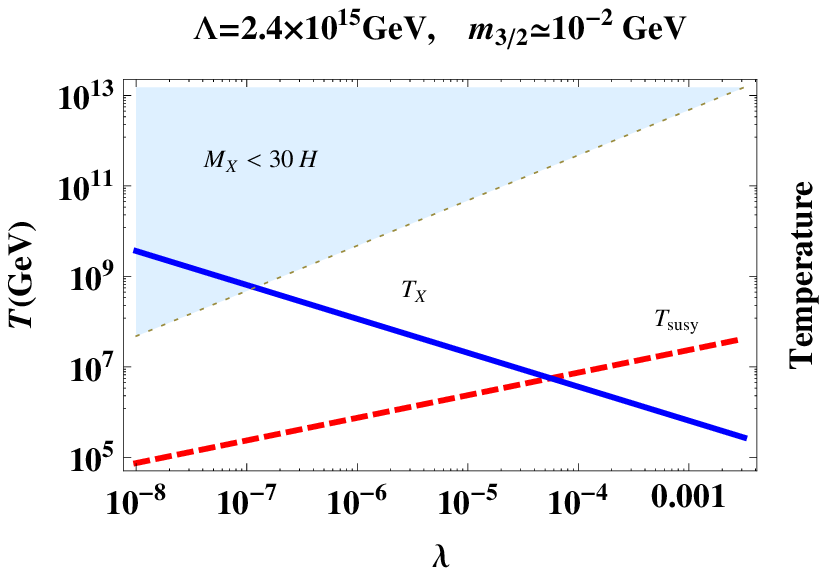} &
{(b2)} \includegraphics [scale=.85, angle=0]{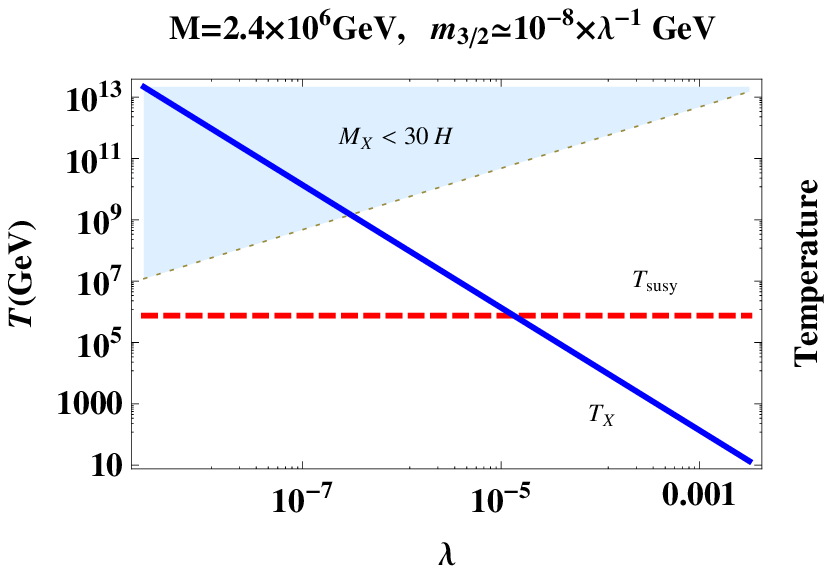}  \\
\end{tabular}
\caption{\small The plots show how the critical temperature $T_\text{susy}$ (red dashed line) of the transition to the susy vacua and the temperature $T_X$ (blue line) of the transition to the metastable vacuum scale with the coupling $\lambda$, for the cases of (a) gravitational stabilization and (b) messenger mass $M\phi\bar{\phi}$. It demonstrates that as the coupling decreases the $T_X$ becomes larger than the $T_\text{susy}$. In the white region there is an efficient damping of the spurion oscillations thanks to a large enough thermal mass i.e. $M_X>30H$. We consider $\lambda=\lambda_\ell=\lambda_q$ and for the messenger mass model, $\Lambda=2.4\times 10^{15}$ GeV.} 
\end{figure}

Decreasing the coupling $\lambda$ also decreases the $T_X$ and generally it cannot get larger that $T_\text{susy}$. Weak $\lambda$ means even weaker stabilization of the spurion i.e. smaller mass. Therefore, models of ordinary gauge mediation with canonical K\"ahler where the susy vacuum is stabilized due to the interactions with the messengers (minimal UV completion) cannot become thermally favourable. 

In the first case of the global limit, (\ref{6}), spurion is stabilized by the corrections in the K\"ahler function. These corrections have also a perturbative quantum origin but they come from the interaction of the spurion with the integrated out degrees of freedom, e.g. with the heavy raifeartons with different coupling $k_o$. So in that case, decreasing the $\lambda$ doesn't change the mass of the spurion.
\section{Conclusions} \normalsize
In this letter we have shown that metastable susy breaking vacua of ordinary gauge mediation with non-vanishing gaugino masses can be thermally selected.
The thermal selection favours a small coupling between the messengers and the spurion and it can be realized for generic initial vevs of the fields. 

The stronger a field is coupled to the thermal plasma the larger is the 'thermal screening' effect on the tree level parameters. Decreasing the coupling makes the zero temperature potential to dominate quickly over the finite temperature corrections. In the case of messengers, SM gauge bosons don't let the thermal mass to drop below $gT$. On the other hand, the spurion being coupled with the coupling $\lambda$ feels only slightly the thermal effects  if $\lambda$ is small enough. Hence, the spurion zero temperature potential can emerge at higher temperatures than the tree level potential of messengers (which is responsible for the tachyonic origin). The conclusion is that the temperature $T_X$ at which the metastable susy breaking vacuum appears can be larger than the critical temperature $T_\text{susy}$ of the transition towards the susy vacuum. This happens in models where the spurion is stabilized due to K\"ahler corrections. Therefore, the spurion zero temperature mass is unaffected by decreasing $\lambda$  because it originates from the interaction with integrated out heavy fields and not from the interaction with messenger fields. A coupling $10^{-8} \lesssim \lambda  \lesssim 10^{-4}$ can make the metastable vacuum thermally favourable for gravitino with ${\cal O}(10^{-3}-1)$ GeV mass.

Let us note that if the messengers have a Yukawa coupling $k$ to SSM fields, e.g. to Higgses like in (\ref{Hq}), and $k>g$ the thermal mass of messengers is further enhanced. This can relax the upper bound on the coupling $\lambda$ for the thermal selection of the metastable vacuum. On the the other hand, we ask for a weakly interacting spurion, which implies that $X$ should not directly couple to any observable field, like in (\ref{HX}), with a coupling $\epsilon>\lambda$.

The selection of the metastable susy breaking vacuum takes place thermally. The oscillations of the spurion are efficiently damped and thus there is no late entropy production. The reheating temperature has to be high enough in order that the messengers to get thermalized and also, the system of fields to get localized in the origin. For small coupling $\lambda$  and temperatures higher than about ${\cal O}(10^7-10^9)$ GeV the thermal selection of the metastable vacuum is realized. It is interesting to note that leptogenesis scenarios can be accommodated in these ordinary gauge mediation models.

We believe that this cosmological constraint on the coupling can be a guide for the hidden sector gauge mediated susy breaking model building.
\section*{Acknowledgments}
\vspace*{.5cm}
\noindent 
ID would like to thank NTU.Athens for hospitality. This work was partially supported by the 
EC 6th FrameworkProgramme MRTN-CT-2006-035863 and by the Polish
Ministry for Science and Education under grant N N202 091839.
    
\end{document}